# Thermally Activated Fracture of Porous Media

Alessio Guarino[1,2] and Sergio Ciliberto[1]

[1] *Laboratoire de Physique de Ecole Normale Supérieure de Lyon, 46 Allée d'Italie, 69364 Lyon, France*

[2] *Université de la Méditerranée Aix-Marseille II, 8 Bd Charles Livon - 13284 Marseille, France*

*Abstract*

The lifetime of a porous media, submitted to a constant subcritical stress, is studied by means of a numerical model. This model is based on a spring network where the porosity is represented by missing springs. The dynamics is produced adding thermal fluctuations in spring forces. The lifetime predicted by the models are compared to experimental data of delayed fracture of porous media submitted to three-point flexion fracture experiments.

## 1. Introduction

When a constant stress is applied to a solid, it will eventually break, even if the applied stress is below the experimental breaking threshold [1-7]. This phenomenon is known as subcritical rupture. The time interval since the application of the stress to the time at which the macroscopic fracture of the sample is observed, i.e. the lifetime of the sample, strongly depends on the temperature, the applied stress and the disorder of the material. In disordered materials several microcracks are observed before the macroscopic failure of the sample (fracture precursors) and thermal activation is believed to be the driving mechanism for microcrak nucleation [7-10]. Nonetheless, it has been shown that thermal fluctuations are too small to activate microcracks in the times measured in the experiments [11,12]. The hypothesis that disorder amplifies the effect of thermal fluctuations has been confirmed both theoretically and experimentally [10-13]. It has been shown that one can define an effective temperature, which is given by the product of the thermal temperature by a term that depends on the disorder distribution. In this paper, we are interested to check whether the same mechanism is observed when disorder is represented by porosity (i.e. a homogenous media with holes) in the 2d spring network model with thermal fluctuations. Indeed, despite their great importance in technological applications, very little attention has been given to the study of delayed rupture of porous media. We show that the lifetime of the samples is strongly affected by the porosity and follows an Arrhenius law similar to that of disordered materials. We also compare the lifetime prediction of our model to experimental data of delayed fracture of porous media.

## *2. The Model*

We model a bidimensional porous elastic media as a square network of springs whose nodes can move only along an axis perpendicular to the undeformed plane of springs. The elastic restoring force of a spring is proportional to the variation in displacement along the moving axis. For simplicity, we set the elastic constant between force and displacement equal to unity. This is one of the most used and simplified models of an antiplane deformation [14-18]. A constant force $F_o$ is applied at the two opposite sides of the lattice, the direction of the force being parallel to the axis of the node movements. Dynamics is introduced in the system by adding to spring forces a fluctuating component which simulates thermal noise. The fluctuations in force $\delta f_j$ that occur in time on spring $j$ are assumed to follow a Gaussian probability distribution with zero mean value and variance $K_BT$ where $K_BT$ represents the thermodynamics temperature in the unit of square force. From a physical point of view, the introduction of the thermal noise in the model is justified by the fact that springs have a damping, which is a source of thermal noise because of the fluctuation dissipation theorem. Whenever the force on a spring exceeds a breaking threshold $f_c$, the spring is cut and, since it is never repaired, the process is irreversible. In a perfectly homogeneous net all the thresholds are equal ($f_c=1$ for each spring). We define $F_c$ as the critical force of the sample, which is the smallest force that can break the sample when $K_BT=0$. The time scale in the simulation is given by the constant time between two configurations of force fluctuations in the system, and the length scale as the distance between two nodes of the lattice.

The aim of this paper is the study the delayed rupture of porous media. We model a porous medium as an elastic homogenous lattice (homogenous spring network) with holes ($f_c=0$). These holes are modeled by initially broken (missing) springs, which represent the porosity of the system. Porosity, $\Phi$, is defined as the ratio between the number $n_a$ of absent springs over the total number of springs $N$, i.e.

$\phi = \frac{n_a}{N}$. The system has a percolation transition at $\Phi=0.5$ [14, 15]. It is important to mention that there is a fundamental difference between "damage" and porosity. Indeed, porosity is randomly distributed, while the distribution of broken springs due to "damage" is leaded by the long range stress (current) re-distribution. This implies that the increase of the porosity and the damage growth are two different processes. This is proved by the fact that the number of springs broken by thermal activation, from the instant in which the sample is loaded until its rupture, presents a strong dependence on temperature. This effect is clearly shown in fig. 7 for various initial porosity. Thus in general the number of broken bonds is not a good parameter to predict the final failure of the sample.

The Young modulus $E$ of the network is defined as the ratio between an imposed force F and the relative (numerically calculated) global displacement. For a homogeneous network, the Young modulus $E_o$ is equal to 1. We find that, in our model, the Young modulus depends linearly on the porosity. Numerical data, figure 1a, suggest that $E = (1-2\Phi)E_o$. Remarkably, this is the same expression that has been proposed on an empirical base by Mc Adam [20] and analytically by Shahidzadeh-Bonn [23], Mackenzie [21] and of Hashin [22]. The same result has been experimentally found in three-point flexion fracture experiment carried on porous bars of size *9 x 80 x 1.6 mm* made of sintered glass beads [23]. The critical force $F_c$ of the network also depends on the porosity $\Phi$. The plot on figure 1b, shows the critical strength $F_c$ as a function of the porosity $\Phi$ for values of the porosity included in the interval [0-35%]. In this interval, numerical data are very well fitted by the exponential function, which agrees with previous theoretical predictions and experimental observations [24].

## *3. Numerical results*

In order to study the delayed rupture of our model, we measure the lifetime $\tau$ of the sample when a subcritical constant force $F_o < F_c$ is applied for different values of the temperature T to a network of porosity $\Phi$. We focused on the study of the network rupture far from the percolation transition because, close to this point, there are practically no fracture precursors, i.e. the failure of a very few springs is enough to break the whole sample, and the process is then trivial. The lifetime $\tau$ of samples with $\Phi=10\%$ are plotted as a function of $\frac{(F_c-F_o)^2}{E}$ for different values of the temperature $T$ in figure 2a), on a semi-log scale,. In the figure squares, circles, triangles, diamonds and crosses represent, respectively, results for simulation with $K_BT=0.001, 0.002, 0.005, 0.01$ and $0.02$. The solid lines are best–fits performed assuming $\tau \approx e^{\alpha\frac{(F_c-F_o)^2}{E}}$. This law has been analytically predicted and checked numerically in similar homogeneous and disordered networks [12, 26-29] and in experiment [28].

The numerical results of fig.2a) clearly show that the lifetime of the samples depends exponentially on $\frac{(F_C-F_O)^2}{E}$. The fitting parameter $\alpha$ has a linear dependence on $1/2K_BT$, as it is shown in figure 2b), i.e. $\tau \approx e^{g\frac{(F_c-F_o)^2}{2EK_BT}}$ and $\alpha = \frac{g}{2K_BT}$. Here, $g$ is equivalent to an activation volume, a parameter that is independent on the temperature. These results are analogous to the homogenous case (network with no holes and no disorder) [10, 26,27] and to the disordered case [29]. In both cases a dependence of $\tau$ on $e^{g\frac{(F_C-F_O)^2}{2EK_BT}}$ has been observed. In figure 3a), we plot $\tau$ as a function of $\frac{(F_C-F_O)^2}{2EK_BT}$ for different values of the sample porosity, at $K_BT=0.01$. The range of the porosity goes from 5% up to 28%. In the figure, circles, triangles, squares and diamonds, represent, respectively, simulations at $\Phi=5\%, 10\%, 20\%$ and $28\%$. The solid lines in fig. 3a are best exponential fits performed using $\tau \approx e^{g(\Phi)\frac{(F_c-F_o)^2}{2EK_BT}}$, where the activation volume $g(\Phi)$ is a fitting parameter independent of $K_BT$. This is shown in Fig.3b) where the values of $g(\Phi)$ computed at various $K_BT$, are plotted as a function of $\Phi$. The figure shows that the activation volume $g(\Phi)$ does not depend on the temperature $K_BT$ and that a linear dependence of $g(\Phi)$ on the porosity can be assumed : $g(\Phi)=\gamma(\chi\Phi+1)$, where $\gamma$ and $\chi$ are constants, that in this case take the values: $\gamma=0.75$ and $\chi=27.6$. The fact that the activation volume increases with porosity can be explained by the fact that voids increase the size of initial defects of the sample. As a consequence the stress concentration on the voids increases too.

The numerical results suggest that $\tau = \tau_o e^{\frac{g(\phi)(F_c-F_o)^2}{2EK_BT}}$. In fig. 4 the ratio $\frac{\tau}{\tau_o}$ is plotted as a function of $\frac{g(\phi)(F_c-F_o)^2}{2EK_BT}$. The figure shows the results of all the runs we performed on 50x50 networks. All points collapse on the solid line corresponding to:

$$\frac{\tau}{\tau_o} = e^{\frac{g(\phi)(F_c-F_o)^2}{2EK_BT}} = e^{\frac{\gamma(F_c-F_o)^2}{2E_oK_BT}\frac{\chi\Phi+1}{1-2\Phi}} \qquad eq.(1).$$

In conclusion, we found that the lifetime of the network strongly depends on the porosity and the imposed load, eq 1. As expected, the lifetime quickly decreases as the imposed load is increased. Eq. 1 also shows that the lifetime rapidly decreases when the porosity is increased[1] and that the argument of the exponential function diverges for Φ=0.5, which is the percolation threshold.

## 4. Life time statistics

The distribution of lifetimes τ obtained in the simulation depends on the porosity; figure 5. For the homogenous network (*Φ=0*), the lifetime distribution is Gaussian. When the porosity is increased (and $F_o$ is unchanged), the average lifetime $<\tau>$ decreases exponentially with a typical width $\frac{\sigma(\tau)}{<\tau>} \approx 0.56 \pm 0.1$, where $\sigma(\tau)$ is the lifetime standard deviation. The lifetime distribution goes from a Gaussian (at low porosity) to a Poissonian (at large porosity), independently from the imposed temperature. As the porosity is increased, the lifetime distribution gets asymmetric on its left because the lifetime cannot be negative and it tends more and more as Poissonian distribution. This is consistent with the fact that at high porosity, only a few springs, broken by thermal, activation are necessary to obtain the macroscopic failure of the sample, fig. 7. Thus a Poissonian-like distribution is expected. On the contrary, as shown on fig. 7, at low porosity, many precursors (spring broken by thermal activation before the sample failure) are necessary to obtain the macroscopic breakdown. This observation justifies the observation of a Gaussian distribution for low porosity

---

[1] One should not forget that $F_c \approx e^{-\nu\Phi}$

## 5. Discussion and comparison with experimental data

The numerical results descibed in the previous sections show that the lifetime of a porous spring network is well approximated by eq. 1. A similar equation has been derived analytically for homogenous spring networks. It has been shown that the lifetime of these networks is well predicted by $\tau = \tau_o \, e^{\frac{a(F_c - F_o)^2}{2EK_BT}}$, where $a$ is a constant which depends on the geometry of the system and the size of a the initial defect [26, 27]. These results have been checked both numerically and experimentally [26-30]. The fact that the lifetimes of porous and homogeneous spring networks are predicted by the same function, can be explained by a simple "homogenization" process. This means that the porous network composed by springs with threshold $f_{cp}$ and holes (springs with $f_c$=0) can be replaced by a homogeneous network composed by springs with $f_{ch} = F_c$, where $F_c$ represents the critical force of the porous network. The lifetime of such a network is then given by eq. (1).

We compare our numerical results with those obtained by Shahidzadeh-Bonn et al. in three-point flexion fracture experiments on porous materials of controlled porosity [23]. In this experiment, porous glass beads samples were submitted to a constant stress until the macroscopic failure of the sample. The lifetime $\tau_{exp}$ of the samples were measured for different values of the porosity and the applied load. The measures have been performed on samples made of sintered glass beads of diameter *104–125 μm* and Young modulus $E_{glass}$=2 10$^{10}$ Pa. The samples were cut into 9 x 80 x 1.6 mm test bars, thus, they can reasonably be considered bidimensional. Porosity is represented by holes of average diameter of 28 *μm*. The range of porosity $\Phi_{exp}$ used in the experiment is *35% < $\Phi_{exp}$ < 50%*. It has been experimentally verified that the Young modulus of the samples $E_{exp}$ is very well approximated by the equation $E_{exp} = (1 - 2\Phi_{exp})E_{glass}$. Remarkably, this is the same law, that we find in the porous spring network presented in *fig 1a*.

In Fig. 6, we plot the lifetime of the samples as a function of $\frac{(F_{C_{exp}} - F_{O_{exp}})^2}{2EK_BT}$ for three different values of the porosity ($\Phi_{exp}$ =36%, 39.7% and 47%) at temperature T=25 °C. Here, $F_{O_{exp}}$ corresponds to the imposed load, which has been varied from $6 \cdot 10^6$

Pa to $3 \cdot 10^7$ Pa. Instead $F_{C_{exp}}$ represents the strength of the sample, which has been extrapolated from the experimental results. The data analysis show that $F_{C_{exp}}$ scales logarithmically with the porosity [24], which is compatible with previous analytical and experimental results. Please note that the dimensions of the samples (18x18x5 mm) used by Liu in [24] allows us to compare our 2d model with his experimental results.

In fig.6 the dotted lines represent the prediction that has been proposed in ref.[25] and used by Shahidzadeh-Bonn et al. to explain their results:

$$\tau_{exp} \approx e^{\frac{\Gamma^3 E_{exp}^2}{K_B T F^4}}, \qquad eq. (2).$$

where the free parameter $\Gamma$ is the energy needed to create an unit surface crack.

The solid line in fig. 6 corresponds to the best fit performed using :

$$\frac{\tau_{exp}}{\tau_o} = e^{\frac{g_{exp}(\phi)(F_{Cexp} - F_{Oexp})^2}{2 E K_B T}}, \qquad eq. (3)$$

where the activation volume $g_{exp}(\phi)$ is a fitting parameter. Because the limited range of the measured $F_{Oexp}$, one can hardly distinguish which model is the best. Indeed the two fits have a comparable accuracy. However, it has been already established [19,31] that the original Pomeau's theory is unable to explain some features observed in similar experiments performed on microcrystals [5], gels [6], fiberglass and chipboards samples. Equation (2) is based on the idea that the rupture of the sample is due to the reversible nucleation of one preexisting defect, which is thermally activated. On one hand, the defect's nucleation cannot be considered a reversible process. Indeed a small crack produced by a fluctuation does not heal and is a favorable starting point for a further growth. On the other hand, experimental observations show that the sample rupture is due to the nucleation and coalescence of several thermally activated microcracks, which is confirmed by the presence of acoustic emissions. Most importantly, the estimated activation energy $U_{act} = \frac{\Gamma^3 E^2}{F_{Oexp}^4}$ of microcrystals, gels, wood and fiberglass is too large with respect to $K_B T$ to justify the observed lifetimes [19,31]. In the case of the porous glass media presented here, the typical[2] activation energy is $U_{act} = 10^{-7} J$, while the experimental value obtained by the fit is $U_{act\,fit} \approx 10^{-22} J$ and $K_B T = 10^{-21} J$. In conclusion, despite the fact that data seem to be in agreement with the predictions of ref [25], this model fails in giving a satisfactory theoretical explication of the experimental observations.

The experimental data; presented in fig. 6, are also in good agreement with functional form of *eq. 3*. The calculated values of the activation volume $g_{exp}(\phi)$ as a function of the porosity $\Phi_{exp}$ are shown in the inset of fig. 6. Despite the poor statistics, it appears that $g_{exp}(\phi)$ has a linear dependence on the porosity of the samples, i.e. $g_{exp}(\phi) \approx \gamma_e (\chi_e \phi_{exp} + 1)$. Where $\gamma_e = 1.5\ 10^{-23}$ m$^3$ and $\chi_e = 3$ are fitting parameters. This is the same result found in our numerical simulations. The calculated activation volume is of the order of *$10^{-24}$ m$^3$*, which leads to an activation length *$V^{1/3} = 10^{-8}$ m*, which is of the same order of magnitude of those usually found in other studies [30]. This length corresponds to the scale of defects, scratches, or cusps usually contained in real solids. The value of activation length given by the model is then physically reasonable. The lifetime distribution observed in the experiments is a Poissonian one. If one considers that the porosity of the samples used in the experiments is quite high ( > 35 %), this is also consistent with the results obtained in our simulations for networks at high porosity.

## *6. Conclusions*

The aim of the paper is the study of delayed rupture in porous media. We modeled a porous sample with an elastic homogenous spring network with holes (missing springs), where the porosity, *Φ*, is defined as the rate of absent springs over the total number of springs. We have shown that the lifetime is well predicted by eq.1. In this equation the only free fitting parameter *g(Φ)*, is a linear functin of the porosity Φ. We have pointed out that this is equivalent to the lifetime of a homogenous spring network where each spring has a critical force *fc$_h$=F$_c$*, where *F$_c$* represents the critical force of the porous network. Finally, we have compared our numerical results with the data obtained in an experiment of fracture performed on sample made of melted glass micro spheres. We have shown that eq.1 predicts quite well the lifetime of the samples. Furthermore the lifetime distributions observed in experiments is consistent with those of our simulations. The fitting parameter $g_{exp}(\Phi)$ obtained from experimental data also depend linearly on the sample porosity $\Phi_{exp}$. However, the experimental data at our disposal are limited; experiments on a larger range of imposed stress and more statistics are necessary in order to make more convincing conclusions. In order to get closer to the experimental framework it would be interesting to generalize the lifetime prediction to the case of a 3D spring network. Indeed, even if the large aspect ratio of the experimental samples legitimizes the 2D approximation, the third dimension could bring a non-negligible volume contribution.

---
[2] Putting E=7·10$^7$ Pa, Γ=10 J/m$^2$

## Figures

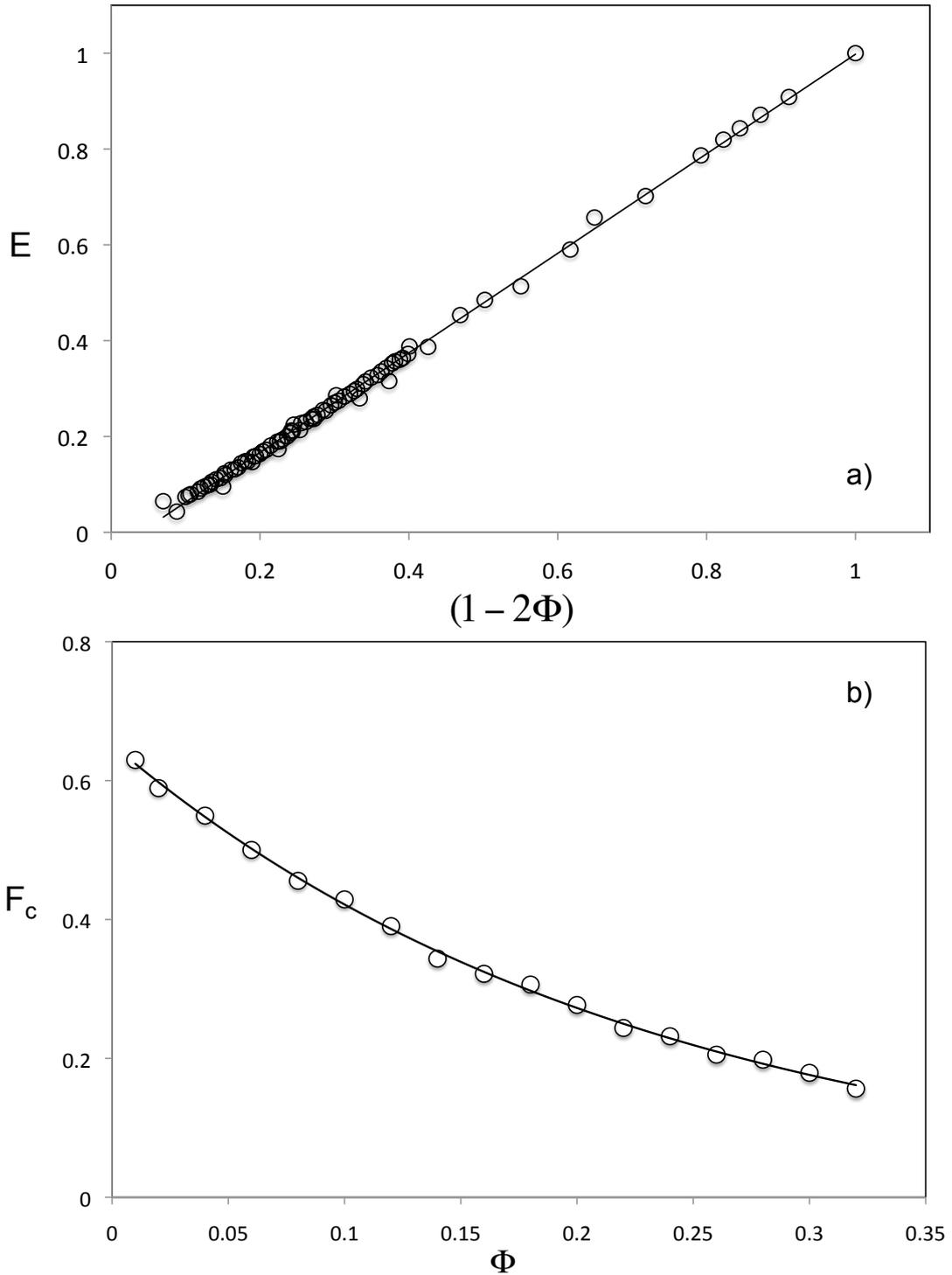

*Figure 1* : a) The Young modulus E of the network as a function of (1- 2 $\Phi$). The solid line represents the bisector, the open circles represent the results of the simulations. b) The critical force of the samples $F_c$ as a function of the porosity $\Phi$. The open circles represents the results of the simulations, the solid line is the best decreasing exponential fit $F_c \approx e^{-\nu\Phi}$. In both figures, each point represent the average on 100 runs performed on 100x100 networks.

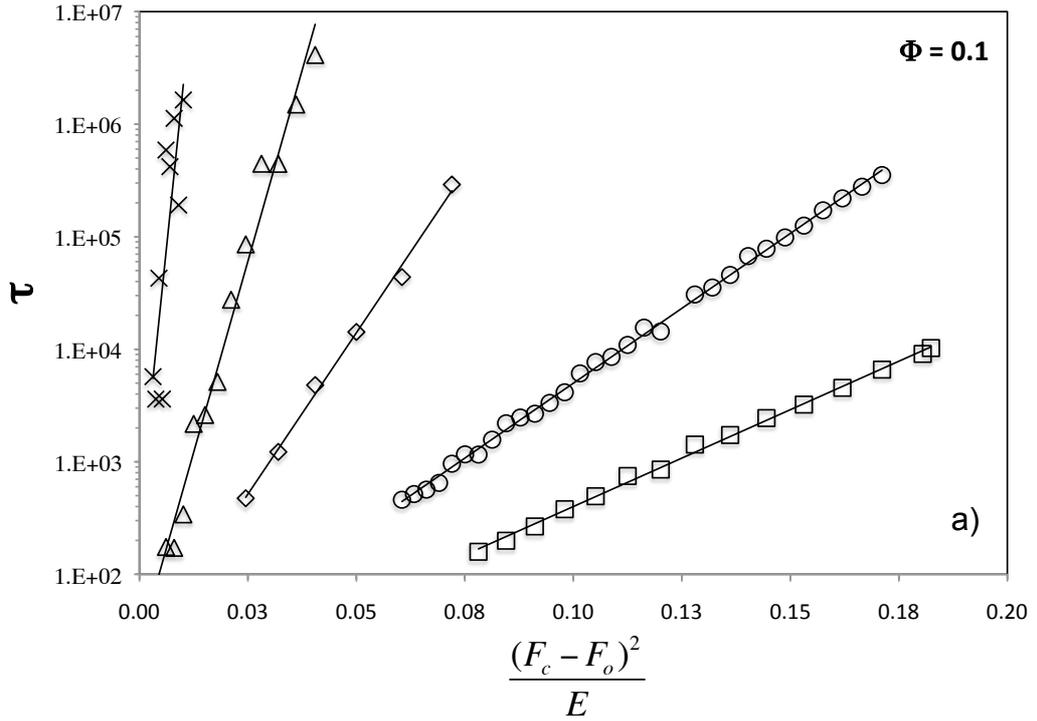

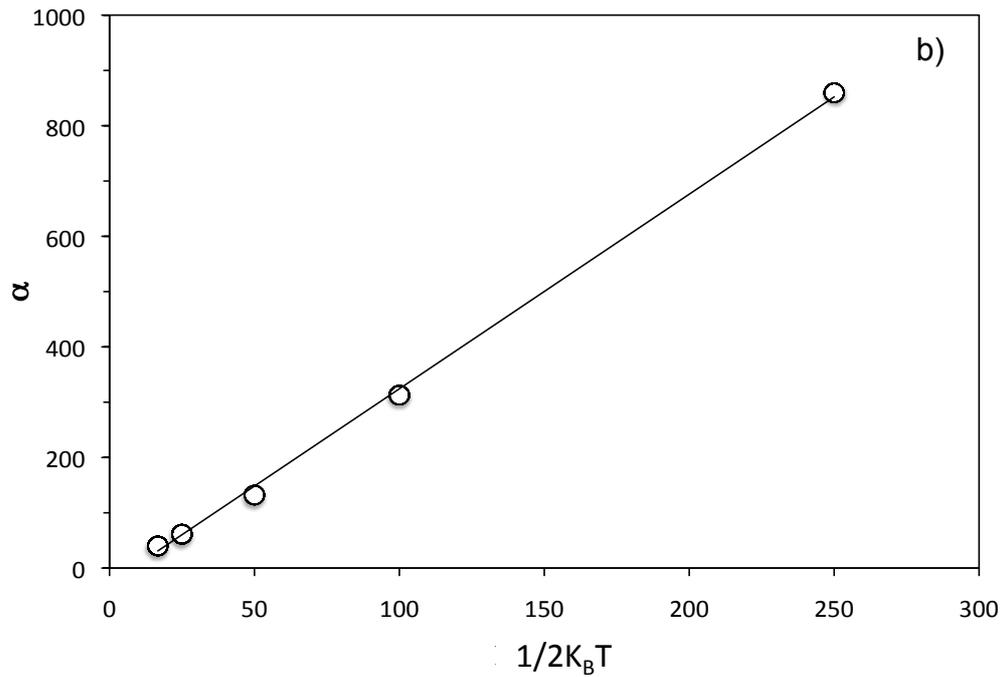

**Figure 2** : a) Lifetime $\tau$ as a function of $\frac{(F_c - F_o)^2}{E}$ for network with $\Phi=10\%$ on a semi-log scale. The solid lines represent the best exponential fit $\tau \approx e^{\frac{\alpha(F_c - F_o)^2}{E}}$. Squares, circles, triangles, diamonds and crosses represent, respectively, results for simulation with $K_BT=0.001$, $K_BT=0.002$, $K_BT=0.005$, $K_BT=0.01$ and $K_BT=0.02$.

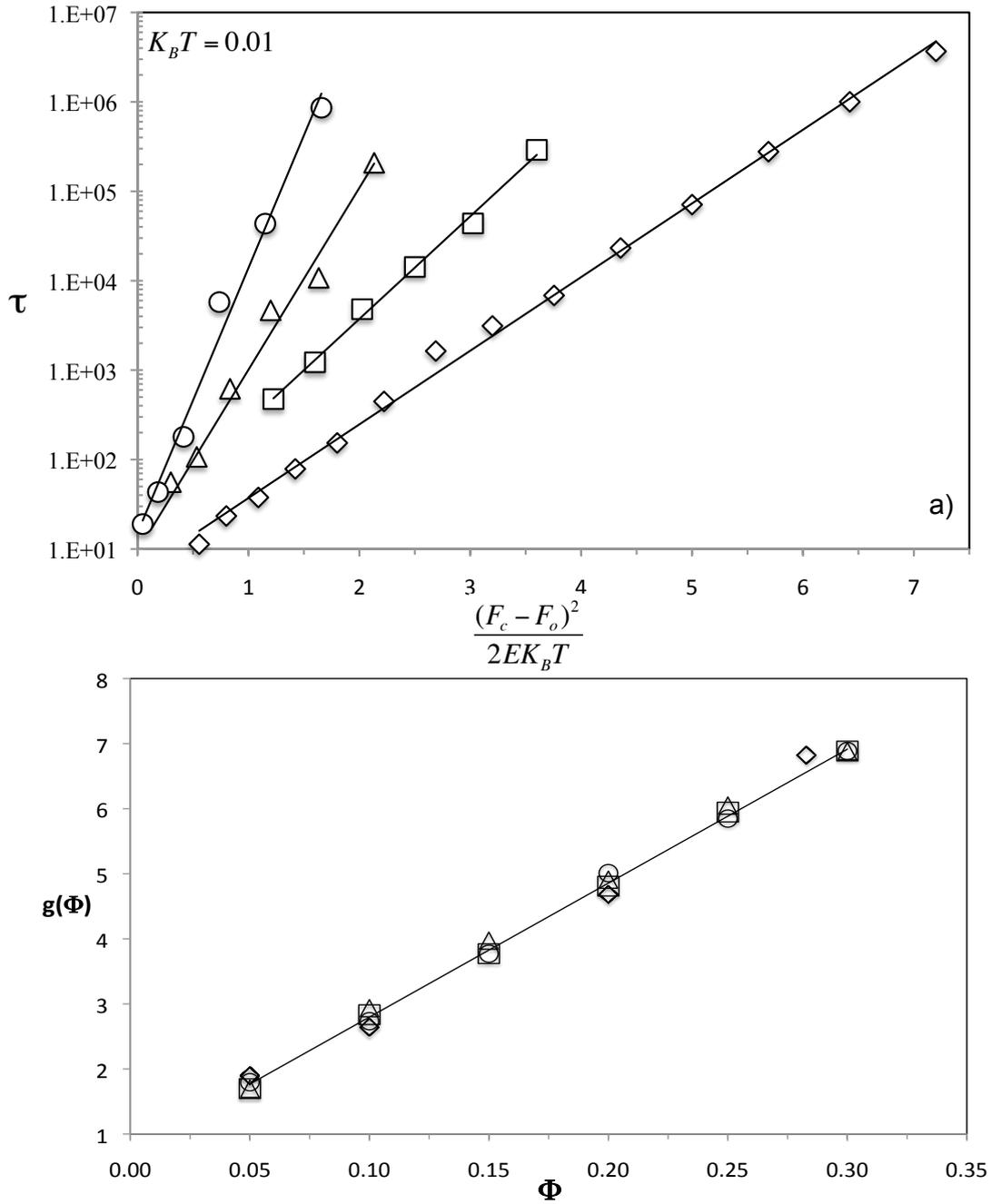

***Figure 3*** *: a) Lifetime $\tau$ as a function of $\frac{(F_C-F_O)^2}{2EK_BT}$ for samples with different porosity $\Phi$ on a semi-log scale . The range of the porosity goes from 5% up to 28 %. Circles, triangles, squares and diamonds, represent, respectively, simulations run with $\Phi=5\%$, $\Phi=10\%$, $\Phi=20\%$ and $\Phi=28\%$. For all run $K_BT=0.01$. The solid lines represent the best exponential fit $\tau \approx e^{g(\Phi)\frac{(F_c-F_o)^2}{2EK_BT}}$, where the activation volume $g(\Phi)$ is the activation volume a fitting parameter. b)The fitting parameter $g(\Phi)$ as a function of the porosity $\Phi$ for $K_BT=0.001$ (circles), $K_BT=0.005$ (diamonds) , $K_BT=0.01$(squares) and $K_BT=0.02$ (triangles).*

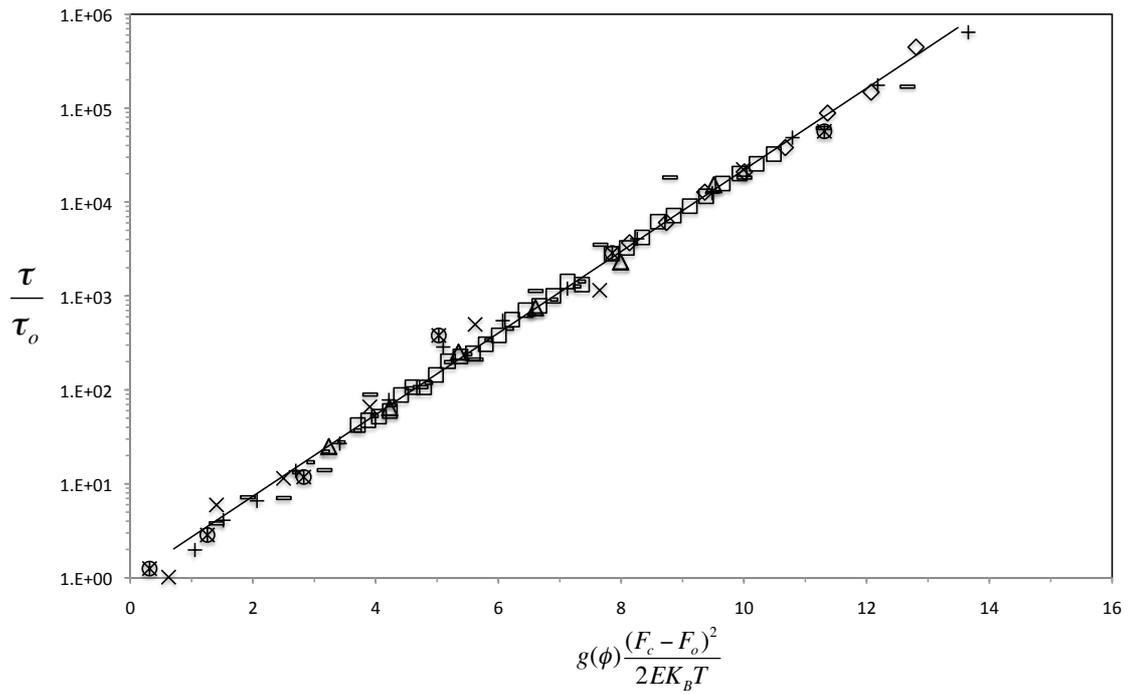

***Figure 4*** *: Lifetime τ as a function of* $\frac{g(\phi)(F_c - F_o)^2}{2EK_BT}$ *on a semi-log scale. All data previously presented are shown.*

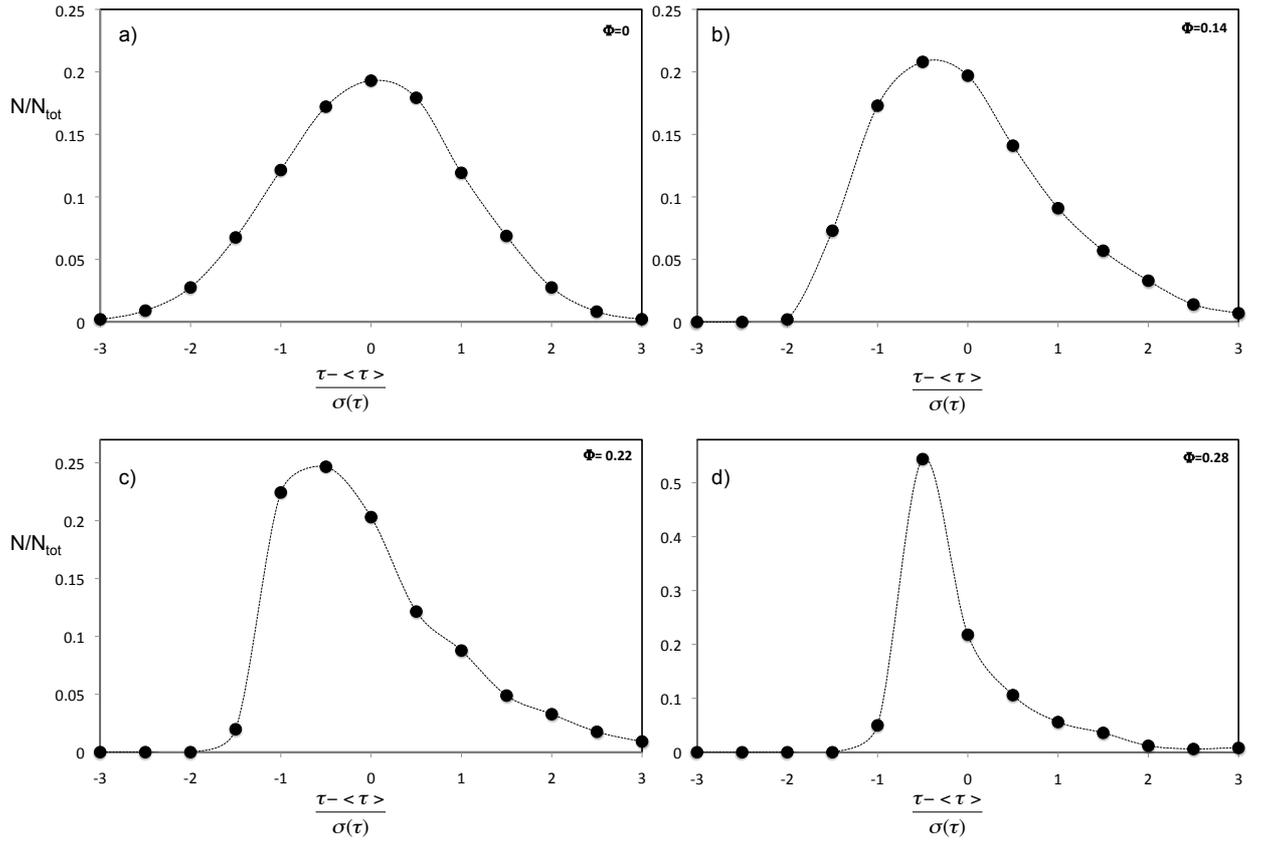

***Figure 5*** *: The lifetime distribution as the function of the normalized variable* $\dfrac{\tau - <\tau>}{\sigma(\tau)}$ *for different porosities : a) Φ=0, b) Φ =14% c) Φ =22% and d) Φ =28%. Here, <τ> and σ(τ) represents, respectively, the average and the standard deviation of τ. For all run $K_BT=0.01$.*

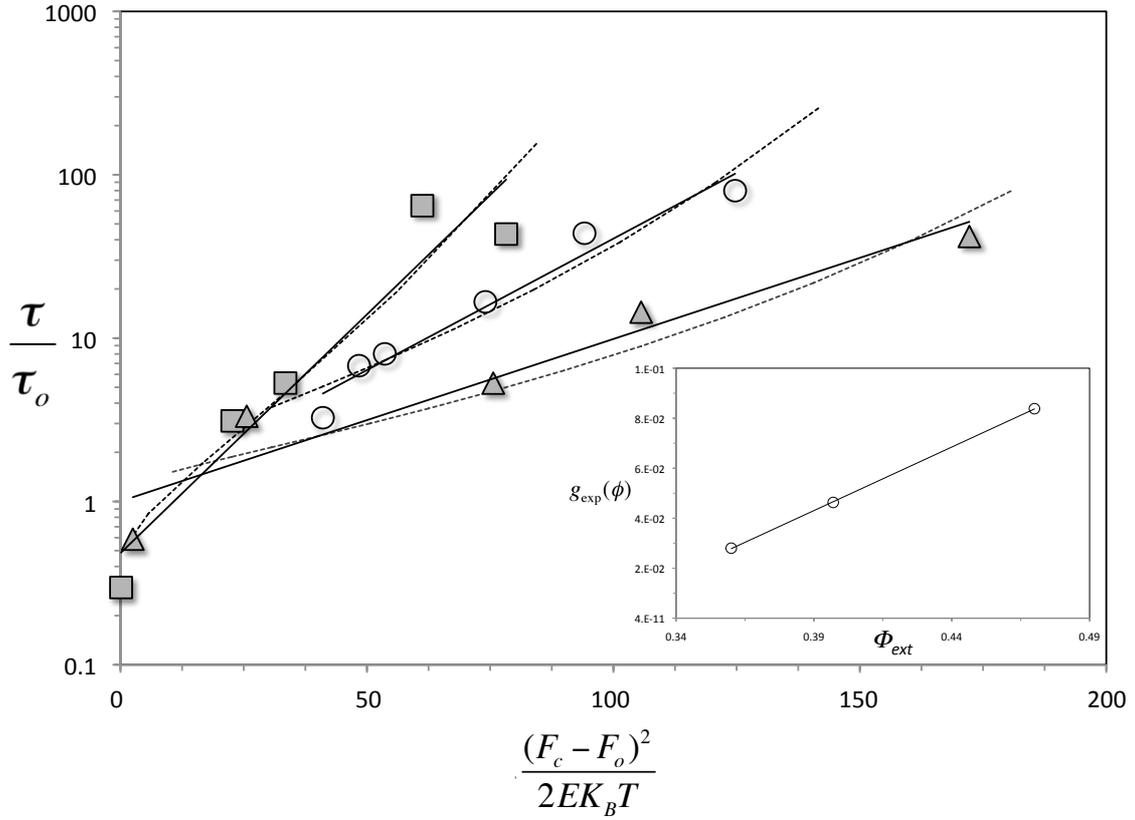

***Figure 6*** : *a) lifetime of the samples as a function of $\frac{(F_c - F_o)^2}{2EKT}$ for three for $\Phi_{exp}$ =36% (squares), 39.7% (circles) and 47% (triangles). Experiments were run at room temperature (T=25 °C). The solid line represents the best fit for $\frac{\tau}{\tau_o} = e^{\frac{g_{\exp}(\phi)(F_c - F_o)^2}{2EKT}}$, where the activation volume $g_{\exp}(\phi)$ is a fitting parameter. The dotted lines represent the prediction $\tau \approx e^{\frac{1}{F^4}}$ that has been proposed by Y. Pomeau. Inset) The the activation volume $g_{\exp}(\phi)$ as a function of the porosity $\Phi_{exp}$. The solid line represents $g_{\exp}(\phi) \approx \gamma \phi_{\exp}$.*

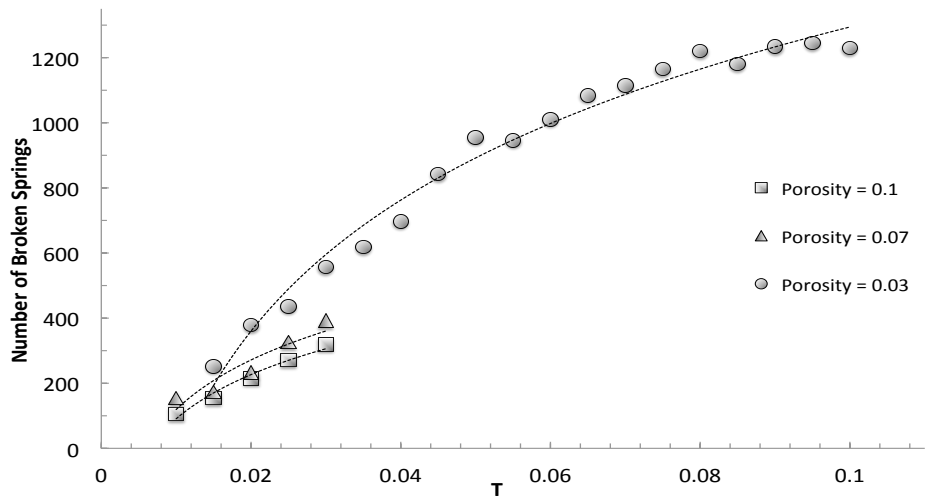

*Figure 7* : *Number of springs broken until rupture by thermal activation as a function of temperature T. Circles, triangles and squares, represent, respectively, simulations run with (Φ,Fo)=(10%,0.15), (Φ,Fo)=(7%,0.15) and (Φ,Fo)=(3%,0.7). Each point represents the average value obtained over 10 simulations.*